\documentclass[notitlepage,aps,pra,onecolumn,superscriptaddress,groupaddress]{revtex4-1}

\usepackage[left=2cm,top=2cm,right=2cm,headsep=0.5cm,headheight=0.5cm,nofoot]{geometry}

\usepackage{mathrsfs,mathpazo,avant,tabularx,units}
\usepackage[utf8]{inputenc}
\usepackage[table]{xcolor}
\usepackage{colortbl}
\usepackage{indentfirst}
\usepackage{stmaryrd}
\usepackage{amssymb,amsmath,amsthm,amsfonts,amsbsy}
\usepackage{bm,bibunits,color,chngcntr,epsfig,epstopdf,graphicx,dsfont}
\usepackage{hyperref,lipsum,,makecell,mathrsfs,rotating}
\usepackage[english]{babel}
\usepackage[normalem]{ulem}

\DeclareMathOperator{\Tr}{Tr}
\usepackage{braket}

\newcommand{\be}{\begin{equation}}
\newcommand{\ee}{\end{equation}}
\newcommand{\bea}{\begin{eqnarray}}
\newcommand{\eea}{\end{eqnarray}}

\newcommand{\I}{\mathds{1}}

\def\C#1{\mathcal #1}

\definecolor{gray}{gray}{0.9}

\begin{document}
\newtheorem{theorem}{Theorem}
\newtheorem{prop}[theorem]{Proposition}
\newtheorem{corollary}[theorem]{Corollary}
\newtheorem{open problem}[theorem]{Open Problem}
\newtheorem{conjecture}[theorem]{Conjecture}
\newtheorem{definition}{Definition}
\newtheorem{remark}{Remark}
\newtheorem{example}{Example}
\newtheorem{task}{Task}

\title{Experimental simulation of quantum superchannels}

\author{Hang Li}
\affiliation{Beijing Academy of Quantum Information Sciences, Beijing 100193, China}
\author{Kai Wang}
\affiliation{CAS Key Laboratory of Theoretical Physics, Institute of Theoretical Physics,
Chinese Academy of Sciences, Beijing 100190, China}
\affiliation{School of Physical Sciences, University of 
Chinese Academy of Sciences, Beijing 100049, China}
\author{Shijie Wei}
\affiliation{Beijing Academy of Quantum Information Sciences, Beijing 100193, China}
\author{Fan Yang}
\affiliation{State Key Laboratory of Low-Dimensional Quantum Physics and Department of Physics, Tsinghua University, Beijing 100084, China}
\author{Xinyu Chen}
\affiliation{State Key Laboratory of Low-Dimensional Quantum Physics and Department of Physics, Tsinghua University, Beijing 100084, China}
\author{Barry C. Sanders}
\affiliation{Institute for Quantum Science and Technology, University of Calgary, Alberta T2N 1N4, Canada}
\author{Dong-Sheng Wang\footnote{wds@itp.ac.cn}}
\affiliation{CAS Key Laboratory of Theoretical Physics, Institute of Theoretical Physics,
Chinese Academy of Sciences, Beijing 100190, China}
\author{Gui-Lu Long\footnote{gllong@tsinghua.edu.cn},}
\affiliation{State Key Laboratory of Low-Dimensional Quantum Physics and Department of Physics, Tsinghua University, Beijing 100084, China}
\affiliation{Beijing Academy of Quantum Information Sciences, Beijing 100193, China}

\date{\today}



\begin{abstract}
Simulating quantum physical processes has been one of the major 
motivations for quantum information science. 
Quantum channels, which are completely positive and trace preserving processes,
are the standard mathematical language to describe quantum evolution, 
while in recent years quantum superchannels have emerged as the substantial extension.
Superchannels capture effects of quantum memory and non-Markovianality more precisely,
and have found broad applications in universal models, algorithm, metrology, discrimination tasks, as examples.  
Here, we report an experimental simulation of qubit superchannels 
in a nuclear magnetic resonance (NMR) system with high accuracy,
based on a recently developed quantum algorithm for superchannel simulation. 
Our algorithm applies to arbitrary target superchannels,
and our experiment shows the high quality of NMR simulators
for near-term usage.  
Our approach can also be adapted to other experimental systems
and demonstrates prospects for more applications of superchannels. 




\end{abstract}

\maketitle

\section{Introduction}

Quantum simulation is one of the original motivations for quantum computing~\cite{Fey82}.
Although being noisy without the aid of quantum error correction~\cite{NC00}, 
experimental quantum simulation is valuable 
for verifying quantum algorithms and protocols,
for developing new quantum information-processing techniques,
and even for the exploration of quantum advantage~\cite{Pre18}. 
General quantum evolution is described as completely positive mappings~\cite{CHOI1975285},
which can describe both unitary and non-unitary processes, 
including measurements. 
Studying non-unitary dissipative processes are important to understand the physics of decoherence~\cite{Bre03},
quantum error correction~\cite{NC00}, and so on.
In recent years, quantum simulation of channels, including open-system dynamics, 
have been studied both theoretically and 
experimentally~\cite{BBW07,PPK+11,WBOS13,WS15,Wang16,SSP14,SSBP15,TV17,TFS+16,CMP+18,LLW+17,XWP+17,MMR18,LXW+18,PJO+20,GRM20}. 

Similar to quantum channels~\cite{CHOI1975285} 
which describe the relationship of input-output states for a quantum system, 
quantum superchannels~\cite{CDP08a,CDP08,CDP09}, also known as supermaps or combs, 
describe the relationships between input and output quantum channels. 
Although a superchannel can also be treated as a channel, 
it captures some peculiar features more precisely, 
such as quantum non-Markovianality~\cite{LHW18} and quantum resources~\cite{RevModPhys.91.025001}.
In recent years, superchannel theory has been widely used 
for studying channel discrimination and quantum metrology~\cite{CAP08,8678741}, 
ebit-assisted quantum communication and error correction~\cite{LB12}, 
a computing model without definite causal order~\cite{CAPV13}, 
quantum von Neumann architecture~\cite{W20_choi,Wang_2022,LWLW23} 
and quantum algorithms like quantum machine learning and quantum optimization~\cite{VPB18,MBW+19,HKP21,W21_model}.


Experimental quantum simulation is indispensable for some applications 
especially when the simulated target is hard to obtain.
Nuclear magnetic resonance (NMR) has been well developed as a sophisticated technology in recent decades for quantum information processing~\cite{NC00,MLSK02,VC05,Xin_2018}.
Different from other platforms~\cite{LJL+10},
liquid-state NMR quantum simulators realize entangling operations on 
the so-called pseudo-pure states and benefit from 
its computer-aided high-fidelity pulse-engineering technology and controlling in full range of the system dynamics.
Although being limited on qubit numbers and sampling cost,
which are similar with some NISQ (noisy intermediate-scale quantum) tasks~\cite{Pre18},
NMR quantum simulators are able to  
simulate quantum systems of small-to-medium sizes
with complex or time-dependent Hamiltonian and test new protocols, 
such as open-system dynamics~\cite{XWP+17}, quantum phase transition~\cite{peng2009quantum},
gate characterization~\cite{lu2015experimental}, measuring correlation functions~\cite{li2017measuring}, quantum imaginary evolution~\cite{cao2022quantum}, heat conduction~\cite{wei2023quantum}
and quantum energy teleportation~\cite{RKM+23}.

In this work, we implement quantum superchannels based on a recent simulation algorithm~\cite{WW23}.
A central part of the algorithm is the decomposition into a convex sum 
of generalized extreme superchannels,
which not only benefits practical simulation, 
but also is relevant for the study of information-theoretic features of channels and superchannels,
e.g., for quantum channel capacity~\cite{Smi08}.
Our theory applies to arbitrary form of superchannels,
and it employs a convex-sum decomposition to reduce the circuit simulation cost.
A unitary circuit that simulates a channel or superchannel 
is further converted to a Hamiltonian evolution for the NMR experimental simulation.
Our 4-qubit NMR simulator, assisted by the simulation algorithm,
is able to realize any qubit superchannel with high fidelity (see Fig.~\ref{gen-ext}). 
Its circuit contains a pair of pre- and post-unitary operators on the input channel 
with ancillary qubits serving as quantum memory. 
We experimentally carried out a few tasks, 
including randomly generating so-called extreme superchannels,
a convex-decomposition of random non-extreme superchannels,
and also random dephasing superchannels.
Our experiment can also be viewed as a first step to confirm the feasibility 
of the recent construction of prototypes of quantum von Neumann architecture~\cite{LWLW23} 
which has a close relation with superchannels.
We also theoretically demonstrate the application of superchannels for
noise-adapted quantum error correction of the amplitude damping channel in the appendix~\ref{app2}.

The remainder of our paper is organised as follows. 
Section~\ref{sec:rep} introduces the algorithm we use simulating superchannels. 
Section~\ref{sec:exp_sim} presents the experimental method and results.
We summarise our paper in Section~\ref{sec:sum}. 
Further numerical details are reported in the Appendix~\ref{sec:ap} and some other examples of superchannel in Appendix~\ref{app2}.  

\section{The algorithm}
\label{sec:rep}

Our goal is to experimentally simulate arbitrary superchannels within a good accuracy.
Usually, quantum evolutions are in general described by completely positive trace-preserving (CPTP) maps,
also known as quantum channels~\cite{NC00}
\be \C E(\rho)=\sum_{i=1}^r K_i \rho K_i^\dagger,\,
\rho \in \C D(\mathscr{H}),\,
\sum_i K_i^\dagger K_i=\I\ee
for~$\{K_i\}$ the Kraus operators~\cite{Kra83}. 
As an example, a unitary `qudit' evolution $U\rho U^\dagger$,
$U\in SU(d)$~\cite{WHSK20}
satisfies $U^\dagger U=UU^\dagger=\I$, with $d=\text{dim}(\mathscr{H})$.

Channel-state duality~\cite{Jam72,CHOI1975285}
maps a channel $\mathcal{E}\in\mathscr{L}(\mathscr{D})$ into a Choi state
\begin{equation}\label{eq:choi}
\omega_\C E:=(\mathcal{E} \otimes \mathds{1}) (|\omega\rangle\langle\omega|),
\end{equation}
with~$|\omega\rangle:=\frac{1}{\sqrt{d}}\sum_{i=0}^{d-1}|i,i\rangle$ a maximally entangled state,
also known as a (generalised) Bell state. 
The rank of the Choi state equals the rank of the channel, which is the minimal number of Kraus operators.

As channels can be viewed as states,  
the operations on them are further defined as superchannels~\cite{CDP08a,CDP08,CDP09}.
Similar with channels, superchannels can also be well represented by quantum circuits,
Kraus operators, and also Choi states. 
Given a channel with a set of Kraus operators $\C E=\{K_i\}$ and Choi state $\omega_{\C E}$, it is changed by a superchannel $\hat{\C S}$ according to 
\be \hat{\C S}(\omega_{\C E})=\sum_a S_a \omega_{\C E} S_a^\dagger\ee
with 
\begin{equation}
S_a= \sum_m K_w^{ma} \otimes K_v^m 
\label{A}
\end{equation}
and $\sum_a S_a^\dagger S_a=\I$ for trace preserving. 
We place a hat on the symbol for superchannels to avoid confusion. 
Subscripts~$v$ and $w$~(\ref{A}) are pre- and post-unitary operators on channels. 
Kraus operators of the output channel are represented as 
\be F_i^a= \sum_m K_w^{ma}K_i K_v^{m,t},\ee
with $K_v^m=\bra{m} V \ket{0}$, $K_w^{ma}=\bra{m} W \ket{a}$, and $t$ stands for transposition.




Given an arbitrary superchannel, 
an algorithm has been developed recently for the task of circuit simulation of the superchannel~\cite{WW23}. 
The algorithm also explores the convexity of the set of superchannels. 
Being convex, there are extreme points that
cannot be written as any convex combination of others~\cite{bengtsson_zyczkowski_2006}.
Choi proved that 
a channel $\C E$ is extreme iff $\{K_i^\dagger K_j\}$ is linearly independent~\cite{CHOI1975285}, 
which has been extended to superchannels~\cite{DPS11}. 
This yields an upper bound on the rank which is a necessary condition,
leading to the notion of generalized extreme points~\cite{Rus07}, or gen-extreme points~\cite{W20_choi}.
Previous studies~\cite{Rus07,WBOS13,WS15,Wang16,W20_choi,WW23} shows that 
decomposition via convex combination of gen-extreme points is a viable approach. 
In this work, we adapt this algorithm to our NMR simulator 
to realize arbitrary qubit superchannels to high accuracy. 

\begin{figure}
    \centering
    \includegraphics[width=0.5\textwidth]{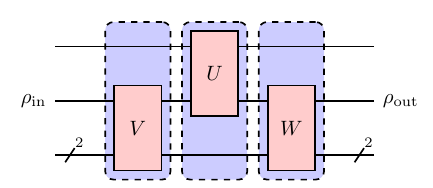}
    \caption{Circuit form of a qubit gen-extreme superchannel}
    \label{gen-ext}
\end{figure}

The circuit form of a qubit gen-extreme superchannel is shown in Fig.\ref{gen-ext}.
It contains 4 qubits and 3 steps for the evolution:
a pre- and post-unitary operation, and the input channel in the middle.
The top register is the ancilla to realize any qubit channel,
and it is proven that a single qubit is enough~\cite{RSW02,WBOS13}.
The 2nd register is the qubit system,
and the remaining two are the ancilla to realize the superchannel.
A direct application of Stinespring dilation would require 3 more qubits,
hence a lot more gates, 
with one (two) for the channel (superchannel),
which can be easily verified from their ranks. 

Our algorithm $\C A$ accepts an arbitrary target superchannel $\hat{\C S}$ as the input, in the form of its Choi state~$\omega_{\hat{\C S}}$, for instance,
and uses an optimization scheme from a built-in package of MATLAB~\cite{WW23} to minimize the trace distance $d(\omega_{\hat{\C S}}, \omega_{\hat{\C S}'})$.
This trace distance represents simulation accuracy
for
\be 
\hat{\C S}' = \sum_{i=1}^{4} p_i \hat{\C S}^{ g}_i,
\label{convex_de}
\ee
which is a convex combination of gen-extreme superchannels
with $p_i$ as a probability
and~$\hat{\C S}^{ g}_i$ as a gen-extreme superchannel. 
Our numerical simulation can guarantee the accuracy in the order of $10^{-3}$ to $ 10^{-4}$~\cite{WW23}.
A gen-extreme superchannel is parameterized based on the 
cosine-sine decomposition of unitary operator~\cite{Wang16}. 

To implement a unitary circuit $U$, the NMR system 
encodes a qubit as the nuclear spin states, 
and converts $U$ into a Hamiltonian evolution $\text{e}^{-\text{i}tH}$, 
which are realized by a sequence of so-called control pulses which manipulates 
the interaction among nuclear spins.
It does not directly decompose $U$ as a sequence of elementary gates,
such as Hadamard and controlled-NOT~\cite{NC00}, and then implement each gate approximately 
which may result in a quick accumulation of errors. 
Instead, it employs algorithms to design control pulses as an engineered Hamiltonian evolution.
Namely,
the gradient ascent pulse-engineering (GRAPE) technique~\cite{khaneja2005optimal,ryan2008liquid} 
is used to design the pulse sequence and achieve the optimal control of radio-frequency field of NMR spectrometer. 
For a given unitary operation $U$, 
the principle of GRAPE is to calculate the gradient of the fitness function corresponding to the forward and backward unitary propagators, 
and the obtained gradient indicates the direction that the control pulses should be optimized to improve the fitness function. 
In general, the GRAPE algorithm is implemented fully on a classical computer. 

To verify the quality of experimental simulation, 
we perform quantum process tomography~\cite{NC00} for randomly chosen input channel $\C E$ 
and output channel $\hat{\C S}(\C E)$ from a superchannel $\hat{\C S}$.
We shall note that for arbitrary random superchannels, 
our algorithm above only works effectively for small dimensions 
as the number of parameters for a superchannel scales exponentially with the physical system size. 
However, we leave it for further study of efficient simulation 
for special types of channels and superchannels of higher dimensions.

This completes the description of our algorithm. 
In the next section, we present the details for the implementation of a few simulation tasks:
random gen-extreme superchannel, random dephasing superchannel,
and also a demonstration of the convex decomposition method. 

\section{Experimental simulation}
\label{sec:exp_sim}

Several superchannels are demonstrated experimentally in the NMR system. We firstly achieve a random gen-extreme superchannel in experiment, 
displaying the accuracy of superchannel simulation in the NMR system.
Then we achieve a dephasing superchannel~\cite{PKS+21}, 
which is the analog of dephasing channels~\cite{NC00} that only affect phase information without the loss of energy.
Finally, we demonstrate the decomposition of a random superchannel, 
showing a great agreement with the theory~\cite{WW23}. 

\begin{figure}
    \includegraphics[width=0.8\textwidth]{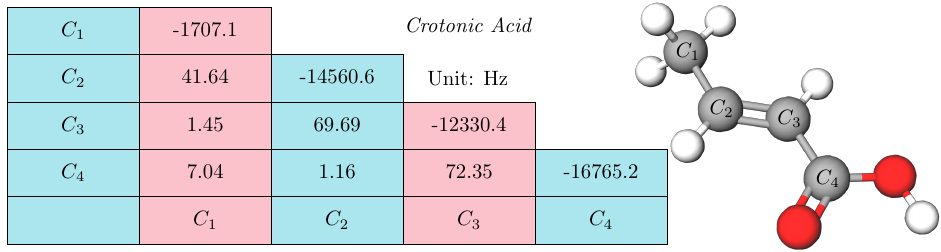}
    \caption{\textbf{Structure and parameters of trans-crotonic acid.} The diagonal and off-diagonal elements in the table are the chemical shifts of spins and J-coupling strengths between spins, respectively.}
    \label{fig:mole_info}
\end{figure}

All experiments are conducted in a liquid NMR system, 
where the sample, $^{13}C$-labeled \textit{trans-crotonic acid} molecules dissolved in \textit{Acetone d6}, 
is placed into a Bruker Avance III 400 MHz spectrometer at the temperature of 303K. 
The molecule contains four carbons, $C_1$, $C_2$, $C_3$, $C_4$, acting as a 4-qubit quantum simulator, whose internal Hamiltonian under the weak-coupling approximation is
\begin{equation}
    H_\text{int} = \sum_{i=1}^{4}\pi\nu_i\sigma_z^i +\sum_{1\le i<j\le 4}^{4}\frac{\pi}{2}J_{i,j}\sigma_z^i\sigma_z^j,
\end{equation}
where~$\sigma_z^i$ and~$\nu_i$ are the Pauli $z$-operator and chemical shift of the $i$-th nuclear spin, respectively ($\nu_1 = -1707.1$ Hz, $\nu_2 = -14560.6$ Hz, $\nu_3 = -12330.4$ Hz, $\nu_4 = -16765.2$ Hz),
and~$J_{i,j}$ is the J-coupling strengths between the $i$-th and $j$-th nuclear spins 
($J_{1,2} = 41.64$ Hz, $J_{1,3} = 1.45$ Hz, $J_{1,4} = 7.04$ Hz, $J_{2,3} = 69.69$ Hz, $J_{2,4} = 1.16$ Hz, $J_{3,4} = 72.35$ Hz). 
The structure and parameters of the \textit{trans-crotonic acid} molecule are illustrated in Fig.~\ref{fig:mole_info}.

For NMR system, the well-established method for initialization is 
to prepare the qubits in the pseudo-pure state (PPS)
\begin{equation}
    \rho_{PPS} = \frac{1-\epsilon}{16}\I^{\otimes4}+\epsilon\ket{0000}\bra{0000},
\end{equation}
where $\I$ is the qubit identity operator and $\epsilon\approx 10^{-5}$ is the polarization~\cite{Xin_2018}. 

A general implementation procedure for simulating the qubit gen-extreme superchannel in Fig.~\ref{gen-ext} with our 4-qubit NMR quantum processor can be achieved as follows.
\begin{enumerate}
\item Preparing $\rho_\text{in}$. The whole system is first  initialized into a 
PPS by using the spatial average technique~\cite{cory1997ensemble}, 
$\rho_{PPS}\simeq \ket{0000}$, starting from the thermal equilibrium state. Then an arbitrary $\rho_\text{in}$ can be prepared afterwards by applying a single-qubit rotation $R_{\phi}(\theta)$ to the work qubit.
\item Constructing superchannel $\hat{\C S}$. This mainly includes applying a pre-operator $V$, an input channel $U$, and a post-operator $W$ sequentially. In general, the three multi-qubit gates can be decomposed into a sequence of single-qubit gates and two-qubit controlled gates with the cosine–sine decomposition (CSD) scheme~\cite{mottonen2004quantum,Wang16}. Here we packed each operator into an individual GRAPE pulse for higher fidelities in experiment, while maintaining the basic structure of superchannels.
\item Measuring $\rho_\text{out}$. Here $\rho_\text{out}$ can be reconstructed through $\rho_\text{out} =\I/2 + \sum_{i\in\{x, y, z\}}c_i\sigma_i$ by performing standard quantum state tomography (QST), where the Pauli components $\sigma_x$ and $\sigma_y$ of $\rho_\text{out}$ can be directly obtained from the spectrum of the work qubit by tracing out the rest qubits, while component $\sigma_z$ can be obtained in the same way by applying a~$\nicefrac\pi2$ rotation readout pulse along the $X$ axis to the work qubit before the measurement.
\end{enumerate}
\noindent
In the following, we show how to select $\rho_\text{in}$ at the first stage of the above procedure, 
to achieve the three operators in the second stage, 
and to characterize the performance of the simulated superchannel with the measured $\rho_\text{out}$ at the last stage in experiment. 

In general, to experimentally determine the dynamics of a channel $\C E$ on an arbitrary single-qubit quantum state $\rho_\text{in}=\tiny{\begin{bmatrix}
    0.5+a & b-ic\\
    b+ic & 0.5-a\\
    \end{bmatrix}}$ ($a$, $b$ and $c$ are all real numbers), preparation and measurement of a quantum state set $\mathcal{B}$ composed of four states are sufficient~\cite{doi:10.1080/09500349708231894}, such that the output state of $\rho_\text{in}$ under $\C E$, $\rho_\text{out}$, can be constructed from the measurement result, $\C E(\mathcal{B})$. In our scheme, we select the quantum state set as $\mathcal{B}=\{\ket{z}, \ket{\bar{z}}, \ket{x}, \ket{y}\}$, where $\ket{z}=\ket{0}$, $\ket{\bar{z}}=\ket{1}$, $\ket{x}=(\ket{0}+\ket{1})/\sqrt{2}$ and $\ket{y}=(\ket{0}+i\ket{1})/\sqrt{2}$, which can form an arbitrary quantum state, pure or mixed, by linear combination. In this case, the output state of an arbitrary quantum state $\rho_\text{in}$ under the quantum channel $\C E$ is
\begin{eqnarray}
\label{eq:state_basis}
    \rho_\text{out}=\C E(\rho_\text{in})&=&(0.5+a-b-c) \C E(\ket{z}\bra{z}) + (0.5-a-b-c) \C E(\ket{\bar{z}}\bra{\bar{z}})\nonumber\\
    &&+ 2b \C E(\ket{x}\bra{x}) +  2c \C E(\ket{y}\bra{y}).
\end{eqnarray}

At the stage of constructing superchannel $\hat{\C S}$, to achieve the optimal control, GRAPE is utilized to pack each of the three unitary operators into one shaped pulse. All shaped pulses are calculated with their fidelities reaching $99.5\%$ and are guaranteed to be robust to the inhomogeneity of radio-frequency pulses.

At the last stage, we exploit a measure of state fidelity for an arbitrary input state $\rho$ under an ideal channel $\C E$ and its experimentally achieved channel $\C E'$. The unattenuated state fidelity $F_{s}$ is 
\begin{equation}
\label{eq:state_fidelity}
F_{s} := \frac{\Tr\left[\C E(\rho)\C E'(\rho)\right]}{\sqrt{\Tr\left[\C E(\rho)^2\right]\Tr\left[\C E'(\rho)^2\right]}},
\end{equation}
where $\C E(\rho)$ and $\C E'(\rho)$ are the ideal and experimental output density matrices corresponding to input $\rho$, and $\C E'(\rho)$.
Thus, $\rho_\text{out}$ in experiment, can be obtained through QST. $F_{s}$ quantifies the similarity of $\C E(\rho)$ and $\C E'(\rho)$ in `direction'~\cite{fortunato2002design}, and is mostly used for measuring the experimental result in the NMR quantum information processing which mitigates the attenuated magnetization caused by dissipation during the unitary operation. 
Similar to the state fidelity definition, the process fidelity between the ideal and experimental realized channel is denoted as
\begin{equation}
    \label{eq:process_fidelity}
    F_p = \frac{\left|\Tr[\chi_\text{exp}\chi_\text{th}^{\dagger}]\right|}{\sqrt{\Tr[\chi_\text{th}\chi_\text{th}^{\dagger}]\Tr[\chi_\text{exp}\chi_\text{exp}^{\dagger}]}},
    \end{equation}
in which $\chi_\text{th}$ and $\chi_\text{exp}$ are the respective ideal and experimentally reconstructed $\chi$ matrix of an arbitrary quantum channel~\cite{zhang2012experimental},
which are equivalent to their Choi states. 

Following the above procedure, 
to change from one experiment to another, 
next we need to select~$\phi$ and~$\theta$ of $R_{\phi}(\theta)$ 
for a specific $\rho_\text{in}$ in set $\mathcal{B}$, 
calculate GRAPE pulses of $\text{pre-}U$ operator, $U$ and post-$U$ operator
for the simulated target superchannel, 
and finally implement the quantum circuit, 
obtaining the experimental results.

\subsection{Simulation of extreme superchannel}

\begin{figure}[t]
    \centering \includegraphics[width=0.7\linewidth]{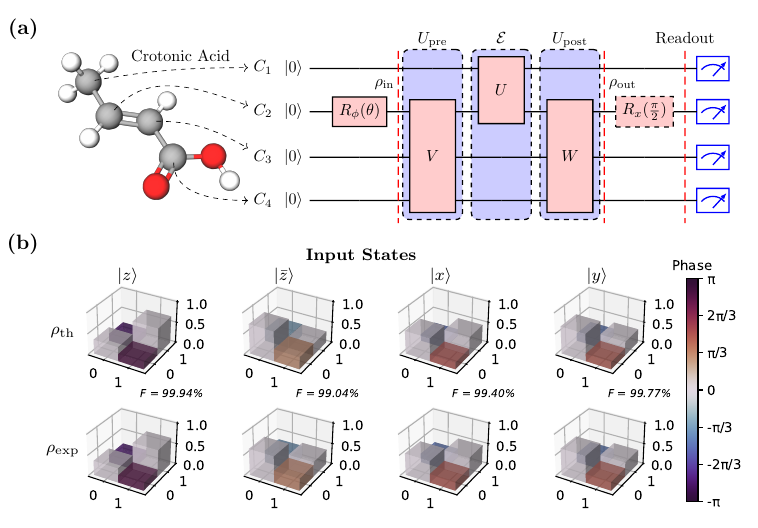}
    \caption{(a)~Quantum circuit for simulating a random extreme superchannel $\hat{\C S}(\C E)$ in experiment. $\rho_\text{in}$ is prepared by applying a rotation $R_{\phi}(\theta)$ to $C_2$, and all components of $\rho_\text{out}$ are obtained by direct measurement or appling a readout pulse $R_x(\frac{\pi}{2})$ before it. (b)~Theoretical (top panel) and experimental (bottom panel) output density matrices of input states $\ket{z}$, $\ket{\bar{z}}$, $\ket{x}$ and $\ket{y}$ under $\hat{\C S}(\C E)$. The amplitude and phase of each entry of density matrices are presented by the height and color of the $3D$ bar, respectively.}
    \label{fig1}
\end{figure}

We experimentally achieve a randomly chosen extreme superchannel $\hat{\C S}$ in the NMR system, as the circuit shown in Fig.~\ref{fig1}(a), where $C_2$ act as the work qubit with $C_1$, $C_3$ and $C_4$ serving as ancillary qubits. Starting from $\ket{0}^{\otimes 4}$, 
a single-qubit rotation $R_{\phi}(\theta)$ is chosen to act on $C_2$ to prepare $\rho_{in}$.
For instance, $R_{\phi}(\theta) = R_{y}(\pi/2)$ can generate $\rho_{in} = \ket{x}$.
An original random channel $\C E$ on the work qubit 
$C_2$ is achieved through a random two-qubit unitary operator $U$ applying to $C_1$ and $C_2$, and then measuring $C_1$ in the final stage, where $C_1$ acts as the ancilla. 
By performing pre operator $V$ and post operator $W$ on $C_2$, $C_3$ and $C_4$, 
where $C_3$ and $C_4$ act as ancilla, and then measuring $C_3$ and $C_4$ at the same time, 
we successfully convert the original channel $\C E$ into $\hat{\C S}(\C E)$. 
The output state $\rho_{out}$ of the work qubit $C_2$ is $\hat{\C S}(\C E)(\rho_{in})$,
which can be further reconstructed with the QST.
In our experiment, the length of GRAPE pulse for implementing $V$, $U$ and $W$ in experiment are 30 ms, 20 ms and 30 ms. More details of the original random channel $\C E$, unitary operators $V$ and $W$ can be found in Appendix~\ref{sec:ap}.

The theoretical and experimental output density matrices under the quantum channel $\hat{\C S}(\C E)$, into which the original quantum channel $\C E$ is converted by the randomly chosen extreme superchannel $\hat{\C S}$, are presented in Fig.~\ref{fig1}(b), where the top (bottom) panel are the output density matrices $\rho_\text{out}$ in theory (experiment) corresponding to the four input bases of $\C B$. The fidelities $F_{s}$ between the theoretical and experimental output density matrices of $\ket{z}$, $\ket{\bar{z}}$, $\ket{x}$ and $\ket{y}$ under the converted channel $\hat{\C S}(\C E)$ are $99.94\%$, $99.04\%$, $99.40\%$ and $99.77\%$, respectively, indicating a very good simulation of the converted channel $\hat{\C S}(\C E)$ in our experiment. For comparison, we also reconstruct the original channel $\C E$, the theoretical and experimental output density matrices under which can be found in Appendix~\ref{sec:ap}.

\begin{figure}
    \includegraphics[width=0.8\textwidth]{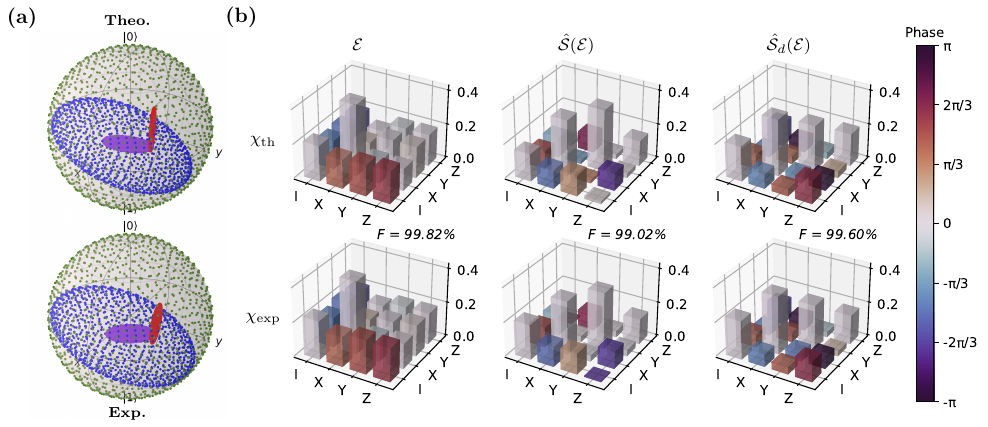}
    \caption{(a)~Input and output states sampled in the Bloch sphere. Theoretical (top panel) and experimental (bottom panel) output states corresponding to the input states (green dots) after a random channel (blue dots), a gen-extreme superchannel (red dots) and a dephasing superchannel (purple dots). (b)~Theoretical (top panel) and experimental (bottom panel) $\chi$ matrices of the random channel $\C E$, converted channels $\hat{\C S}(\C E)$ and $\hat{\C S}_{d}(\C E)$.}
    \label{fig:chi_bloch}
\end{figure}

To generalize our simulation result of the randomly chosen superchannel $\hat{\C S}$, 1000 input states are sampled on the Bloch sphere surface (green dots in Fig.~\ref{fig:chi_bloch}(a)) based on the spherical Fibonacci lattice method, of which the output states can be reconstructed by the measured output states of the four states~(\ref{eq:state_basis}). For comparison, the theoretical and experimental output states of the original random channel $\C E$ are presented as the blue dots in Fig.~\ref{fig:chi_bloch}(a), while that of $\hat{\C S}(\C E)$ are plotted as the red dots. The experimental results of $\C E$ and $\hat{\C S}(\C E)$ are in good agreement with their theoretical ones, and the significant difference between the output results of $\C E$ and $\hat{\C S}(\C E)$ indicates the success of converting the original channel $\C E$ into $\hat{\C S}(\C E)$ by superchannel $\hat{\C S}$.

To fully characterize the original channel $\C E$ and the converted channel $\hat{\C S}(\C E)$, quantum process tomography (QPT)~\cite{NC00} was conducted, and $\chi_\text{exp}$ matrices of both channels were reconstructed from the experimental QST results of our selected input state set $\mathcal{B}$. Thereafter they were transformed into the standard basis set \{I, X, Y, Z\}, as shown in Fig.~\ref{fig:chi_bloch}(b), which reveals further evidence of successfully converting the original channel $\C E$ into $\hat{\C S}(\C E)$ by superchannel $\hat{\C S}$ -- dramatic difference in the amplitude and phase of entries of $\C E$ and $\hat{\C S}(\C E)$.
The process fidelities of the ideal channel $\C E$ and the experimentally achieved channel $\hat{\C S}(\C E)$ are $99.82\%$ and $99.02\%$, respectively, which implies a good verification of accurate channel simulation of $\C E$ and $\hat{\C S}(\C E)$, and further proving the accurate simulation of superchannel $\hat{\C S}$.

\subsection{Simulation of dephasing superchannel}



The dephasing superchannel is also gen-extreme~\cite{PKS+21}. 
In a fixed basis, it is defined to preserve the diagonal elements of the input Choi states while suffering a phase noise which alters the non-diagonal elements.
Actually,
if we consider dephasing channels acting on $d^2$-dimensional systems, then it is easy to obtain that a dephasing superchannel is of the form of dephasing channels.

\begin{figure}
    \centering \includegraphics[width=0.7\linewidth]{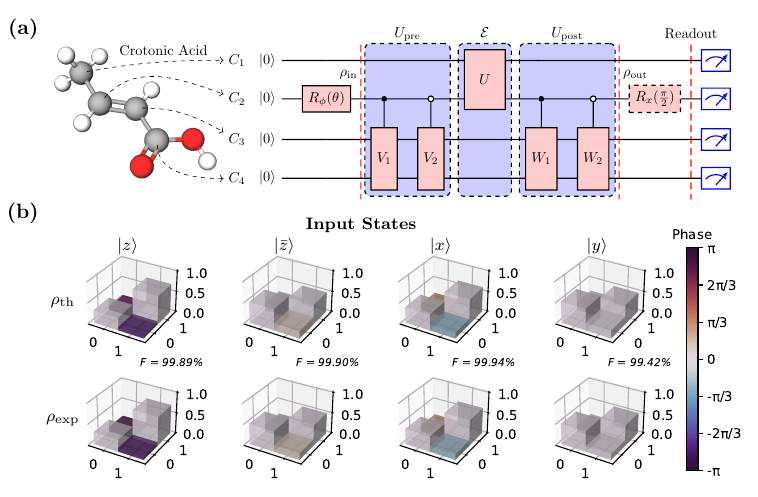}
    \caption{(a)~Quantum circuit for simulating a random extreme superchannel $\hat{\C S}_d(\C E)$ in experiment. (b)~Theoretical (top panel) and experimental (bottom panel) output density matrices of input states $\ket{z}$, $\ket{\bar{z}}$, $\ket{x}$ and $\ket{y}$ under $\hat{\C S}_d(\C E)$.}
    \label{fig2}
\end{figure}

The experimental circuit is shown in Fig.~\ref{fig2}(a).
With similar method with the previous experiment, 
here we use controlled-$V_i$ and controlled-$W_i$ (targetting at $C_3$ and $C_4$ with $C_2$ as the control qubit, and $i\in\{1, 2\}$) as the pre and post operators to 
construct a dephasing superchannel.
as the input channel only acts on the control unit, the diagonal elements of the input Choi states stays the same. 
More details of unitary operators $V_i$ and $W_i$ ($i\in\{1, 2\}$) can be found in Appendix~\ref{sec:ap}.

The whole experiment proceeds as before, the random channel $\C E$ is preserved, i.e., keeping $U$ unchanged, while the controlled-$V_i$s and controlled-$W_i$s are packed into two individual GRAPE pulses with length of 22 ms and 25 ms, respectively. The same state set $\mathcal{B}$ is selected as the input of our simulated channel, and the output density matrices are presented in Fig.~\ref{fig2}(b). The state fidelities $F_{s}$ between the theoretical and experimentally reconstructed density matrices corresponding to the four input states under the channel $\hat{S}_{d}(\C E)$ are $99.89\%$, $99.90\%$, $99.94\%$ and $99.42\%$, respectively. 
Besides, to characterize the function of $\hat{\C S}_d$, one way is to compare the output state of an arbitrary input state before and after the application of dephasing superchannel $\hat{\C S}_d$. 

The sampled theoretical and experimental output states of the converted channel $\hat{\C S}_d(\C E)$ are presented as the purple dots in Fig.~\ref{fig:chi_bloch}(a), which shows the experimental results of $\hat{\C S}_d(\C E)$ are in good agreement with their theoretical ones, and successfully converting the original channel $\C E$ (blue dots) into another channel $\hat{\C S}_d(\C E)$ by a random dephasing superchannel $\hat{\C S}_d$. An alternative way is to reconstruct the process matrices $\chi$ before and after applying $\hat{\C S}_d$, as shown in Fig.~\ref{fig:chi_bloch}(b). The process fidelity $F_p$ between the ideal and experimentally reconstructed $\chi$ matrices of the converted channel $\hat{\C S}_d(\C E)$ achieves $99.60\%$. In a nutshell, both facts indicate a very good simulation of the dephasing superchannel $\hat{\C S}_d$ in our experiment.

\begin{figure}
    \includegraphics[width=0.7\textwidth]{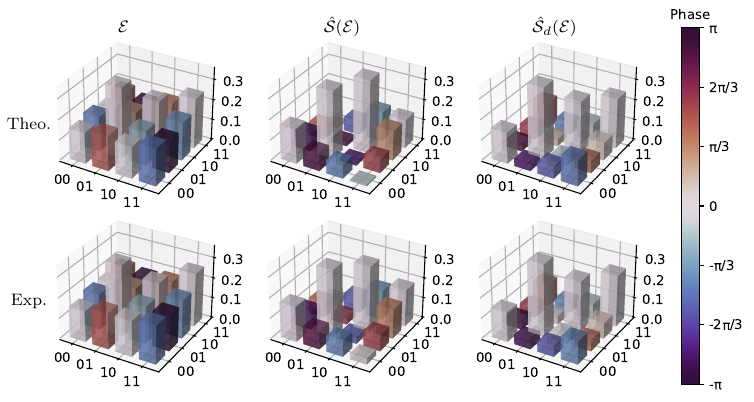}
    \caption{Choi state matrices of channel $\C E$, $\hat{S}(\C E)$ and $\hat{S}_{d}(\C E)$. The top (bottom) panel shows the respective Choi state matrix reconstructed from the theoretical (experimental) Choi state matrices of the random channel $\C E$, $\hat{S}(\C E)$ and $\hat{S}_{d}(\C E)$.}
    \label{fig:choi_state}
\end{figure}

For comparison, the ideal and experimental Choi states of channels $\C E$, $\hat{\C S}(\C E)$ and $\hat{\C S}_{d}(\C E)$ are reconstructed based on the $\chi$ matrices, and presented in Fig.~\ref{fig:choi_state}. 
It shows the dephasing superchannel $\hat{\C S}_{d}$ causes the dephasing of Choi state $\omega_{\C E}$ of the random channel $\C E$ — compressing the amplitude of non-diagonal elements while leaving the diagonal elements unchanged. However, a random superchannel $\hat{\C S}$ (see subfigures in the middle of Fig.~\ref{fig:choi_state}) does not have that feature in general.

\subsection{Demonstration of superchannel decomposition}

\begin{figure}
    \centering \includegraphics[width=1\textwidth]{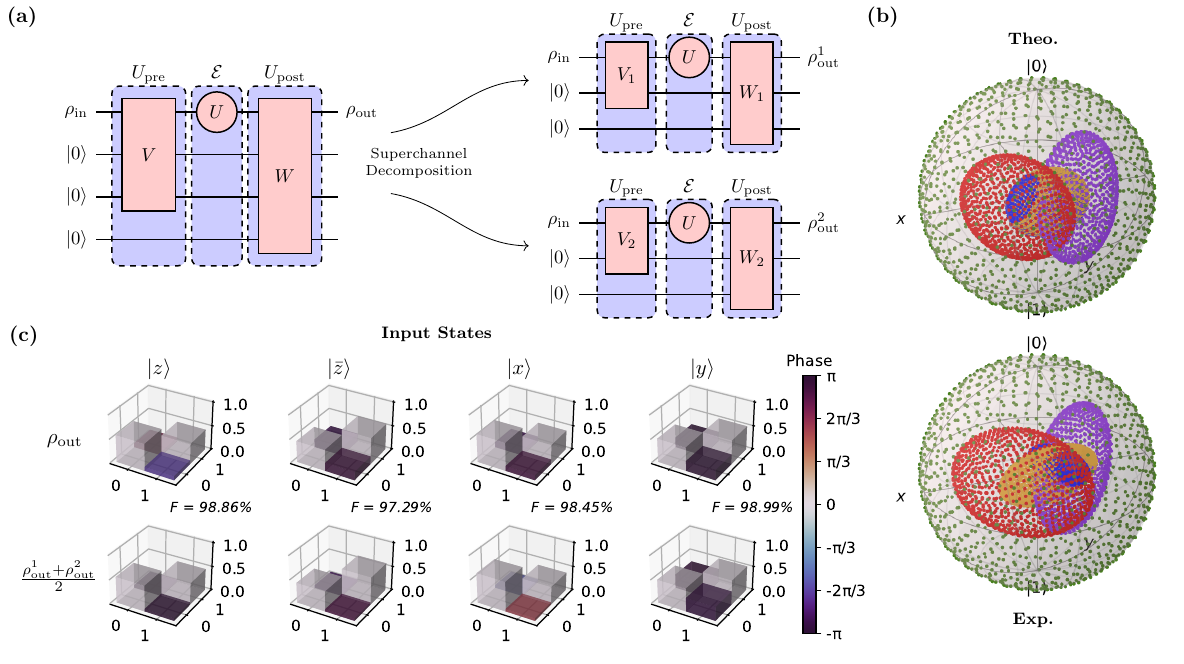}
    \caption{(a)~Superchannel convex-decomposition scheme. A random superchannel simulated in 4 qubits can be decomposed into two 3-qubit extreme superchannles. (b)~Theoretical (top panel) and experimental (bottom panel) output states in the Bloch sphere. The green, blue, red, purple and orange dots represent the sampled input states $\rho_\text{in}$, corresponding output states $\rho_\text{out}$, $\rho_\text{out}^1$, $\rho_\text{out}^2$, and $(\rho_\text{out}^1+\rho_\text{out}^2)/2$, respectively. (c) Experimental output density matrices in our superchannel convex-decomposition scheme. (Top panel) The output states $\rho_\text{out}$ of $\hat{\C S_g}$ in $\ket{z}$, $\ket{\bar{z}}$, $\ket{x}$ and $\ket{y}$ basis, while the bottom panel shows the average of the output states under its decomposed two superchannels, $(\rho_\text{out}^1+\rho_\text{out}^2)/2$.}
    \label{fig:decom}
\end{figure}

We experimentally demonstrate the decomposition of a general superchannel $\hat{\C S_g} $. Restricted by our experimental apparatus, we choose the input channel as an unitary operator $U$ and design our quantum circuit by constructing a random non-extreme superchannel with randomly chosen $V$ and $W$, as shown in the left panel of Fig.~\ref{fig:decom}(a). We demonstrate that this 4-qubit-composed superchannel can be decomposed into two 3-qubit-composed extreme superchannels. Here, we choose $V_i$ and $W_i$ using composition with equal $p_i$~(\ref{convex_de}). Therefore, for an arbitrary input channel $U$ and an arbitrary input state $\rho_\text{in}$, the output state under the general superchannel $\hat{\C S_g} $, which is $\rho_\text{out}$ can be approximated as the average of the output states $\rho_\text{out}^1$ and $\rho_\text{out}^2$ under the two extreme superchannels $\hat{\C S_g^1} $ and $\hat{\C S_g^2} $, i.e., $\rho_\text{out} \approx (\rho_\text{out}^1+\rho_\text{out}^2)/2$.

In our scheme, three groups of experiments corresponding to the circuits in Fig.~\ref{fig:decom}(a) were conducted, where $U$, $V$ and $W$ are generated randomly, while two pairs of $V_i$ and $W_i$ ($i\in\{1, 2\}$) are calculated based on $V$ and $W$. See Appendix~\ref{sec:ap} for more details of unitary operators $U$, $V$, $W$, $V_i$ and $W_i$ ($i\in\{1, 2\}$). 
We pack each unitary operator $V$, $U$, $W$ of the three circuits into an individual GRAPE pulse. The input states $\rho_{in}$ are selected from the input state set $\mathcal{B}$, after which the QST process on $C_1$ is performed to reconstruct the output state $\rho_{out}$, $\rho_{out}^1$ and $\rho_{out}^2$. The theoretical and experimental output states are illustrated in Appendix~\ref{sec:ap}. The state fidelities $F_{s}$ obtained range  from $97.28\%$ to $99.96\%$, where the lower fidelities mainly come from the 4-qubit-composed superchannel circuit whose unitary operators $V$ and $W$ are more complicated, thus the total pulse length of the circuit from preparing pseudo-pure state to making measurements reaches 165 ms, causing larger decoherence effect of $T_2$.

As before, the theoretical and experimental output states of $1000$ input states, which are sampled based on the spherical Fibonacci lattice, under the original superchannel $\hat{\C S_g} $ and the two extreme superchannels $\hat{\C S_g^1} $ and $\hat{\C S_g^2} $ presented in Fig.~\ref{fig:decom}(b), separately. The output states of the sampled input states under the general superchannel $\hat{\C S_g} $ are illustrated in blue dots, while that under extreme superchannels $\hat{\C S_g^1} $ and $\hat{\C S_g^2} $ are illustrated in red and purple dots. The theoretical and experimental output states are in good agreement (the comparatively bad agreement of blue dots corresponding to simulating $\hat{\C S_g} $ can be attributed to the decoherence effect of $T_2$), indicating a well simulation of each superchannel in experiment.

To demonstrate the implementation of the convex-decomposition of a random superchannel $\hat{\C S_g}$ in experiment, we reconstruct the averaged states of the experimental outputs of our selected inputs under the two extreme superchannels $\hat{\C S_g^1}$ and $\hat{\C S_g^2}$, and compare them with their corresponding experimental output states under the original superchannel $\hat{\C S_g}$, as shown in Fig.~\ref{fig:decom}(c). Besides, we calculated the fidelities between the corresponding states in the top panel and the bottom panel of Fig.~\ref{fig:decom}(c), which are $98.86\%$, $97.29\%$, $98.45\%$ and $98.99\%$, respectively. Furthermore, the theoretical and experimental averaged output states of $1000$ sampled input states corresponding to the two extreme superchannels $\hat{\C S_g^1}$ and $\hat{\C S_g^2}$ are presented in orange dots in Fig.~\ref{fig:decom}(b), as a comparison with the original output states (blue dots). The mismatch of them in experiment (blue and orange dots in the bottom panel of Fig.~\ref{fig:decom}(b)) is mainly concentrated on the data in the Bloch sphere along the positive $x$-axis and negative $z$-axis, which are dominated by our imperfect experimental realization of the original superchannel $\hat{\C S_g}$.


\section{Summary}
\label{sec:sum}

In this paper, we experimentally realized an algorithm-assisted NMR simulator of qubit superchannels.
We demonstrate our simulator with three simulation tasks, 
including a random extreme superchannel, 
a dephasing superchannel and a superchannel convex-decomposition scheme.
Furthermore, our experimental result also shows that the superchannel achieved by convex-decomposition has a higher fidelity than that achieved directly, 
on account of the less number of qubits involved in the circuit realization.
The experimentally simulated superchannels are in great agreement with their theoretical counterparts. 
Our experiment verifies the feasibility of the convex channel-decomposition algorithm, 
implying the promising usage of it for higher-dimensional cases and other tasks.


\appendix
\section{Details of experiments}
\label{sec:ap}

\subsection{Random channel} 
\begin{figure}
    \includegraphics[width=0.7\textwidth]{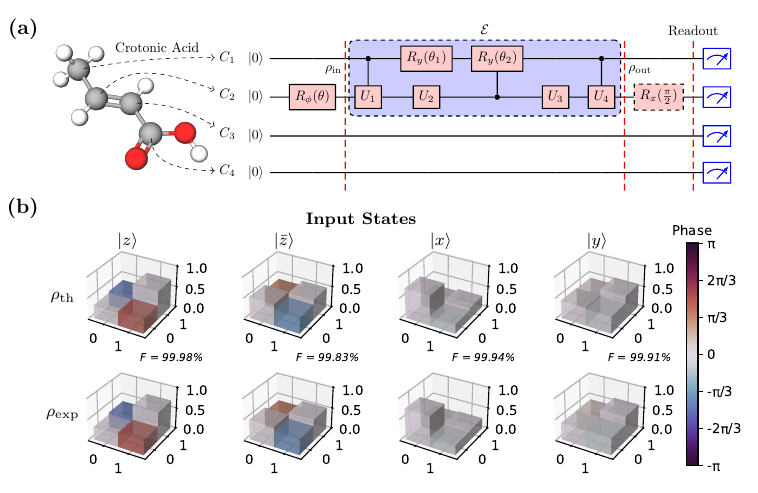}
    \caption{(a)~Quantum circuit for simulating a random extreme superchannel $\C E$ in experiment. (b)~Theoretical (top panel) and experimental (bottom panel) output density matrices of input states $\ket{z}$, $\ket{\bar{z}}$, $\ket{x}$ and $\ket{y}$ under $\C E$.}
    \label{fig:rand_chan_cir}
\end{figure}

In our scheme, a random channel $\C E$ is implemented by a two-qubit unitary with one qubit as the work qubit and the other as an ancilla, and a random chosen $U$ can be found~(\ref{eq:U}). A circuit for simulating a random channel with our $4$-qubit NMR quantum processor can be found in Fig.~\ref{fig:rand_chan_cir}(a), where~$C_2$ serves as the work qubit and $C_1$ as the ancilla. The entire system is initialized in the pseudo-pure state $\ket{0000}$ from the thermal equilibrium state with the spatial average technique at first, and $U$ can be decomposed into a sequence of single-qubit gates and two-qubit controlled gates with the cosine–sine decomposition (CSD) scheme~\cite{mottonen2004quantum}.
\begin{equation}
\label{eq:U}
    U = 
    \begin{bmatrix}
        -0.0109 + 0.1787i & -0.2558 - 0.1492i & -0.2519 + 0.2656i & -0.4561 - 0.7336i\\
        -0.6709 - 0.2262i &  0.1270 - 0.1671i & 0.4623 - 0.3549i & -0.3017 - 0.1551i\\
        -0.1406 - 0.5049i & -0.2872 - 0.5841i & -0.4014 + 0.1515i & -0.0718 + 0.3353i\\
        0.3717 - 0.2321i & -0.6624 + 0.0766i &  0.1922 - 0.5526i & 0.0420 - 0.1389i
    \end{bmatrix}.
\end{equation}

The output density matrices of the input state bases under the random channel $\C E$ in theory and experiment are presented in Fig.~\ref{fig:rand_chan_cir}(b). The state fidelities between the theoretical and experimental density matrices of $\ket{z}$, $\ket{\bar{z}}$, $\ket{x}$ and $\ket{y}$ under the channel $\C E$ are $99.98\%$, $99.83\%$, $99.94\%$ and $99.91\%$, respectively.

\subsection{Extreme superchannel} 
To simulate a random extreme superchannel, two random unitary pre-$U$ (V) and post-$U$ (W) operations of the superchannel $\hat{\C S}$ are generated as follows, 

\begin{equation}
V = {\tiny\begin{bmatrix}
    0.1310 + 0.0036i&0.1140 - 0.2525i&-0.0068 + 0.1055i&-0.0556 - 0.1594i&-0.1454 - 0.1854i& 0.0073 - 0.3811i& 0.2214 - 0.0904i& 0.3471 - 0.6985i\\
    -0.1604 - 0.2878i&-0.1567 - 0.0025i&0.0613 - 0.1076i&-0.2527 + 0.1187i&-0.0739 - 0.7163i& 0.1182 - 0.0921i& 0.2490 - 0.3120i&-0.0706 + 0.2611i\\
    0.0732 + 0.3537i&-0.4630 + 0.0267i&0.0920 + 0.0445i&0.2267 + 0.2262i& 0.1608 - 0.0286i& 0.5365 + 0.2132i&-0.0708 - 0.3283i&-0.0228 - 0.2614i\\
    -0.0093 + 0.1815i&0.1717 + 0.0019i&0.6431 - 0.1916i&-0.2676 - 0.0393i&0.2835 - 0.0428i&-0.2191 - 0.0210i&-0.4000 - 0.2678i& 0.2271 + 0.0038i\\
    0.0826 - 0.2314i&-0.4620 - 0.2645i&0.5516 + 0.0022i&0.2165 - 0.1446i&-0.3134 + 0.0563i& 0.0363 - 0.0046i& 0.0473 + 0.3653i& 0.1578 + 0.1447i\\
    -0.1508 - 0.4899i&0.3662 - 0.0245i&0.1410 + 0.0025i&-0.2153 - 0.0342i&-0.1672 + 0.2146i& 0.5844 + 0.2191i&-0.1869 - 0.0348i&-0.0823 - 0.1694i\\
    0.2754 + 0.0592i&-0.3809 - 0.1076i&-0.3006 + 0.1213i&-0.6761 - 0.1579i&-0.0537 + 0.1575i& 0.1289 - 0.1106i&-0.2538 + 0.0173i& 0.1640 + 0.1696i\\
    -0.2048 + 0.5161i&0.0955 + 0.2651i&0.2211 + 0.1859i&-0.3256 - 0.1205i&-0.2508 - 0.2259i& 0.1189 + 0.1293i& 0.1400 + 0.4275i&-0.2195 - 0.1139i   
\end{bmatrix}},
\end{equation}

\begin{equation}
W = {\tiny\begin{bmatrix}
    0.4438 + 0.3052i&-0.1516 - 0.1924i& 0.1702 - 0.2479i& 0.1763 + 0.1283i&0.0741 + 0.1495i&-0.0073 - 0.2209i&-0.5178 + 0.0332i&-0.4073 - 0.0107i\\
 0.2915 + 0.1024i& 0.1375 - 0.0168i&-0.3881 - 0.2651i& 0.0306 + 0.0003i&-0.1106 + 0.3559i& 0.2050 + 0.3621i&-0.1741 + 0.0693i& 0.5075 - 0.2425i\\
-0.0480 - 0.2797i& 0.0909 - 0.0858i& 0.1075 - 0.1673i&-0.3236 - 0.0472i&0.1355 + 0.3724i&-0.3431 - 0.5753i&-0.1038 - 0.1457i& 0.3381 - 0.0721i\\
-0.3029 + 0.1556i& 0.1901 + 0.4494i&-0.0179 - 0.1726i& 0.0388 - 0.2170i&0.0461 - 0.3691i& 0.1803 - 0.2864i&-0.4075 + 0.3044i& 0.1226 - 0.2016i\\
 0.1913 + 0.3040i&-0.1178 + 0.5217i& 0.2078 - 0.0749i&-0.2228 - 0.3476i&0.3159 + 0.2880i&-0.0994 + 0.1536i& 0.3143 - 0.0292i&-0.1557 - 0.1599i\\
0.3786 + 0.0292i&-0.1600 + 0.0010i& 0.3856 + 0.4572i&-0.3605 + 0.1393i&-0.0270 - 0.2707i&-0.0489 + 0.1073i&-0.1928 + 0.2379i& 0.3690 - 0.0730i\\
-0.0113 + 0.0191i& 0.5498 + 0.1459i& 0.2502 + 0.3443i& 0.4872 + 0.1597i&0.2423 + 0.3037i&-0.1340 + 0.0581i&-0.0583 + 0.1419i& 0.0926 + 0.1658i\\
-0.0737 + 0.3725i&-0.1613 - 0.0762i&-0.0488 - 0.1704i&-0.0245 + 0.4611i&0.3602 - 0.0189i& 0.2744 - 0.2661i& 0.3682 + 0.2291i& 0.2555 + 0.2231i
\end{bmatrix}}.
\end{equation}

\subsection{Dephasing superchannel} 

To simulate a random dephasing superchannel, random unitary pre-$U$ and post-$U$ operations of the superchannel $\hat{\C S}_d$, controlled-$V_i$ and controlled-$W_i$ ($i\in\{1, 2\}$), are generated. The generated $V_i$ and $W_i$ are as follows,

\begin{equation}
V_1 = \begin{bmatrix}
    0.2987 + 0.2302i& 0.4874 - 0.2877i& 0.1190 + 0.2665i&-0.6694 + 0.0645i\\
    0.0898 + 0.2401i&-0.5247 - 0.5788i&-0.3600 + 0.2082i&-0.0291 - 0.3876i\\
    0.7734 - 0.1597i& 0.1368 + 0.1863i&-0.5275 - 0.1207i& 0.1707 + 0.0293i\\
    -0.4084 + 0.0403i&-0.0644 + 0.1087i&-0.6699 + 0.0164i&-0.3253 + 0.5107i
\end{bmatrix},
\end{equation}

\begin{equation}
V_2 = \begin{bmatrix}
    -0.5089 + 0.0961i&-0.3067 + 0.4412i& 0.0754 + 0.5513i&-0.2788 - 0.2359i\\
    -0.1870 - 0.2108i& 0.2685 + 0.6255i&-0.5304 - 0.2773i& 0.0053 + 0.3145i\\
    0.5061 + 0.0705i&-0.0278 - 0.1007i&-0.4042 + 0.3316i&-0.6458 + 0.1939i\\
    0.5896 - 0.2091i& 0.1698 + 0.4562i& 0.0943 + 0.2234i& 0.2692 - 0.4904i
\end{bmatrix},
\end{equation}

\begin{equation}
W_1 = \begin{bmatrix}
    -0.2919 + 0.0605i&-0.6294 - 0.6274i& 0.1046 - 0.0290i& 0.3088 + 0.1191i\\
    0.2956 - 0.0233i&-0.0973 - 0.4107i&-0.3067 - 0.1554i&-0.3989 - 0.6758i\\
    -0.3639 + 0.7948i& 0.1587 - 0.0780i&-0.2993 + 0.2414i&-0.2318 + 0.0552i\\
    -0.0806 - 0.2295i& 0.0100 + 0.0259i&-0.1371 + 0.8387i& 0.3037 - 0.3545i
\end{bmatrix},
\end{equation}

\begin{equation}
W_2 = \begin{bmatrix}
    -0.6174 + 0.3184i& 0.4179 - 0.3318i& 0.2918 + 0.0792i& 0.3384 + 0.1635i\\
-0.5162 - 0.4553i&-0.5212 - 0.1814i& 0.2151 - 0.1841i&-0.0083 - 0.3761i\\
-0.1403 + 0.0197i&-0.4111 - 0.4270i&-0.5477 + 0.1773i&-0.0179 + 0.5448i\\
0.1005 + 0.1163i&-0.1773 - 0.1668i& 0.6157 + 0.3434i&-0.6002 + 0.2446i
\end{bmatrix}.
\end{equation}

\subsection{Superchannel convex-decomposition} 
To simulate a general superchannel $\hat{\C S}_g$ and its decomposed superchannels $\hat{\C S}_g^1$ and $\hat{\C S}_g^2$ in Fig.~\ref{fig:decom}(a), the input unitary channel $U$, pre-$U$ operator V, and post-$W$ are generated randomly as follows. Then two pairs of pre-$U$ operator $V_i$ and post-$U$ operator $W_i$ corresponding to each decomposed superchannel are calculated ($i\in\{1, 2\}$).

\begin{equation}
U = {\footnotesize 
\begin{bmatrix}
    0.6196 + 0.2891i&0.5199 - 0.5120i\\
  -0.7191 + 0.1236i&0.1266 - 0.6720i
\end{bmatrix} } ,
\end{equation}

\begin{equation}
V = {\tiny
\begin{bmatrix}
-0.3048+0.1792i & -0.1955+0.1514i & -0.1071+0.4419i & -0.2905+0.4894i & 0.0582-0.0069i & 0.3937+0.0449i & 0.1629-0.0733i & 0.2968+0.0526i\\
-0.2444+0.2554i & 0.3402-0.3981i & -0.38-0.0044i & -0.0067+0.1559i & -0.3934-0.1094i & -0.2565-0.1723i & -0.2516+0.1297i & 0.2637-0.1417i\\
-0.0471-0.3504i & 0.1682-0.1016i & -0.4623-0.0185i & 0.3695+0.2124i & 0.4099+0.1691i & 0.1721+0.1661i & -0.086-0.2099i & -0.0726-0.3607i\\
-0.3144+0.1616i & -0.2029+0.0234i & -0.2241-0.1805i & 0.2492-0.0861i & -0.3302+0.2253i & -0.0083-0.1936i & 0.3955-0.4893i & -0.2484+0.1617i\\
0.1437+0.323i & 0.1932+0.2012i & -0.1963+0.0867i & -0.2034+0.0288i & 0.1726-0.4514i & -0.3652+0.2439i & 0.3478-0.1881i & -0.2604-0.2417i\\
0.3028+0.1824i & 0.0162+0.1912i & -0.36-0.0257i & 0.4639-0.1851i & -0.053-0.21i & 0.2402+0.1343i & 0.2099+0.2835i & 0.3502+0.2979i\\
-0.3514+0.0389i & -0.0504-0.2043i & 0.375-0.0656i & 0.225-0.191i & 0.0744-0.1392i & -0.1365+0.5133i & 0.0526-0.2765i & 0.4555-0.07i\\
-0.0019+0.3535i & 0.5216-0.3915i & 0.1607-0.0696i & -0.112-0.0962i & 0.3056+0.2732i & 0.3194-0.0075i & 0.279+0.0544i & -0.1099+0.1849i\\
\end{bmatrix}},
\end{equation}

\begin{equation}
\begin{split}
W = {\tiny \left[ \begin{matrix}
    0.2496+0.0149i & 0.2583+0.094i & -0.142-0.1203i & 0.0349-0.3243i & -0.0593-0.2491i & 0.1259-0.1864i & 0.2037-0.0822i & -0.0785+0.048i\\
-0.2373+0.0788i & -0.3037+0.126i & -0.0402+0.1706i & -0.0718+0.23i & -0.1582+0.1418i & 0.1646-0.1858i & 0.1247-0.1802i & -0.2847+0.1219i\\
-0.2248+0.1095i & 0.1179-0.1226i & 0.1822+0.0558i & -0.0741+0.2566i & -0.1138-0.1522i & -0.0315-0.2075i & 0.2547-0.1522i & 0.0425-0.2004i\\
-0.1901+0.1623i & 0.3132+0.0882i & 0.0661-0.1666i & -0.284+0.0335i & 0.0413+0.2787i & -0.1569+0.0769i & -0.1572+0.1256i & -0.0615+0.1315i\\
-0.0658-0.2412i & -0.0596+0.248i & 0.0867-0.3673i & -0.237+0.2021i & -0.0141+0.0435i & -0.0651+0.3i & 0.0707-0.0175i & 0.4078-0.1094i\\
0.1712-0.1822i & 0.1187+0.1043i & 0.3132+0.1139i & -0.1399+0.1597i & 0.2434+0.0886i & 0.3244-0.0918i & 0.0191+0.1291i & -0.1523-0.1812i\\
-0.2493-0.0189i & 0.2004+0.2046i & -0.0233+0.191i & 0.2898-0.1628i & -0.1203-0.0997i & 0.1576+0.2563i & -0.1431+0.1529i & 0.0233-0.1447i\\
-0.2374-0.0784i & 0.1884+0.0826i & 0.1236+0.1263i & 0.2088+0.0682i & 0.0835-0.1652i & -0.1811-0.1144i & 0.2168-0.2539i & 0.2162+0.1115i\\
-0.1046-0.2271i & 0.0901-0.1779i & -0.2164-0.0362i & 0.0399-0.2029i & 0.1211-0.0554i & -0.3369-0.084i & -0.0279+0.3642i & -0.116+0.0211i\\
-0.1484-0.2012i & 0.1407-0.0525i & -0.2549-0.0569i & 0.2266-0.0998i & -0.1081+0.474i & 0.0232+0.0294i & 0.2458-0.2511i & 0.0353+0.0316i\\
-0.2351+0.0851i & 0.1065-0.1222i & -0.0901+0.2486i & -0.0477+0.2347i & 0.2255-0.0802i & -0.0224-0.174i & -0.1716+0.1242i & 0.237+0.0908i\\
-0.2241-0.1109i & -0.266-0.1549i & -0.0406+0.1548i & 0.2583-0.0425i & -0.179+0.1142i & -0.0059-0.0271i & -0.0212+0.3362i & 0.0442+0.0525i\\
0.0403-0.2467i & 0.1307+0.0725i & 0.1723+0.2049i & 0.2006-0.0199i & 0.1575-0.1724i & 0.038+0.4326i & 0.0355-0.1262i & -0.2927+0.3575i\\
-0.0077-0.2499i & -0.2569+0.1936i & -0.0661+0.122i & -0.0947-0.1935i & 0.1183-0.2758i & -0.0135-0.1952i & -0.1709-0.2207i & 0.0604-0.3126i\\
0.2175+0.1234i & -0.053+0.1559i & -0.23+0.2018i & 0.0436+0.1896i & -0.3336-0.0763i & 0.1976-0.1153i & -0.1811+0.0747i & 0.1895+0.2625i\\
0.2498-0.0112i & 0.3396-0.1248i & -0.0057+0.3846i & -0.1327+0.0958i & -0.164-0.0312i & -0.0961+0.1644i & 0.0943-0.1998i & 0.0498-0.1012i\\
\end{matrix}\right.}\\
{\tiny\left.\begin{matrix}
-0.0784-0.2447i & -0.1059-0.187i & 0.2683+0.1565i & 0.027-0.3735i & 0.0801-0.182i & 0.0589+0.1512i & 0.1614-0.0585i & -0.1959+0.2558i\\
0.1948+0.2293i & 0.048-0.3343i & 0.0701-0.1766i & -0.0728-0.1856i & -0.0705-0.2182i & -0.2751+0.0986i & 0.2089+0.1523i & 0.0045+0.0682i\\
0.0825-0.1661i & -0.2516+0.2799i & -0.1183+0.2518i & -0.2201+0.0661i & 0.1242-0.2592i & -0.22+0.0525i & -0.2873-0.0038i & -0.2517-0.0691i\\
-0.157-0.0736i & 0.022+0.1112i & -0.235-0.3061i & 0.1081-0.2763i & 0.1078+0.0399i & -0.1089-0.0951i & -0.0276+0.2065i & -0.333+0.2794i\\
-0.0824-0.1228i & -0.0549-0.1076i & 0.2991+0.014i & -0.0186+0.1768i & -0.0352-0.052i & -0.3052+0.1329i & 0.266-0.0877i & -0.0168+0.0436i\\
0.1974+0.17i & -0.148+0.161i & 0.1659+0.0188i & -0.0182+0.2201i & -0.105-0.0469i & 0.2984-0.1923i & 0.1901+0.1128i & -0.1552+0.3129i\\
0.1318-0.3185i & 0.1564+0.0239i & 0.0877-0.1062i & -0.4498-0.0141i & -0.0473-0.0824i & -0.0954-0.1962i & 0.0109+0.2583i & 0.1798+0.0812i\\
-0.2362+0.2337i & 0.2573-0.0928i & 0.0188-0.1261i & -0.1145+0.1495i & 0.3622+0.0737i & 0.3076+0.0484i & 0.2261+0.1468i & -0.1411-0.0428i\\
0.323+0.1483i & -0.1323-0.1395i & 0.0935-0.0912i & -0.029+0.3175i & 0.1167-0.0644i & -0.1258+0.3647i & -0.1169+0.1682i & -0.0955+0.1469i\\
0.0226+0.0146i & -0.2957-0.083i & -0.0928+0.3398i & 0.0924+0.0761i & -0.0457+0.2595i & -0.0453-0.2416i & 0.0365+0.2071i & 0.0346+0.1146i\\
-0.1721-0.1751i & 0.0594-0.4234i & 0.1996+0.1861i & 0.205+0.0541i & -0.3586-0.0757i & 0.0632-0.169i & -0.2057+0.036i & -0.1029+0.0554i\\
-0.1026-0.0207i & 0.0667+0.1958i & -0.0253+0.092i & -0.0918-0.067i & -0.0848+0.0525i & -0.0045-0.0126i & 0.3188-0.4419i & -0.4276+0.1375i\\
-0.2824+0.1499i & -0.1682+0.024i & -0.059-0.0036i & 0.1132+0.0792i & -0.1176-0.2134i & -0.2738+0.0064i & -0.1411-0.127i & -0.0674+0.0204i\\
-0.1578-0.0919i & 0.0187-0.0196i & -0.5032+0.017i & 0.0682+0.0755i & -0.1524+0.0931i & -0.1336+0.0936i & 0.0804+0.1285i & -0.0352+0.2828i\\
-0.0483-0.0613i & -0.1126+0.1125i & 0.0355-0.063i & 0.2123+0.3211i & 0.4206-0.0092i & -0.1389-0.0905i & -0.0502+0.016i & 0.0641+0.2646i\\
0.3464-0.0108i & 0.3327-0.0229i & 0.0219-0.0296i & 0.0952-0.0145i & -0.1348+0.3624i & -0.2264+0.1113i & 0.0436-0.1429i & -0.1608+0.0548i\\
\end{matrix} \right] },
\end{split}
\end{equation}

\begin{equation}
V_1 = { \tiny 
\begin{bmatrix}
-0.4371-0.2427i & -0.2991+0.0099i & -0.543-0.2371i & -0.5449+0.1117i\\
-0.4979+0.0455i & -0.0157+0.2448i & 0.4424-0.5705i & 0.2641+0.3145i\\
-0.3409-0.3658i & -0.2866-0.3124i & -0.213+0.2186i & 0.6782-0.1312i\\
-0.4157-0.2778i & 0.8157+0.0725i & -0.024+0.1845i & -0.0852-0.1938i\\
\end{bmatrix}},
\end{equation}

\begin{equation}
W_1 = {\tiny\begin{bmatrix}
0.1311+0.3283i & 0.1787+0.1497i & 0.2948-0.0795i & -0.1267+0.144i & -0.2625+0.1616i & 0.435+0.4329i & 0.348+0.2178i & -0.118+0.1909i\\
-0.3088-0.1721i & -0.0507-0.2383i & 0.4115-0.0159i & 0.3146-0.33i & -0.3464+0.4922i & 0.0009+0.0816i & -0.1953-0.1556i & 0.0655+0.0522i\\
0.2655-0.2335i & -0.4459-0.1607i & -0.0586+0.491i & 0.0964+0.1417i & 0.2059+0.2077i & 0.3649+0.1429i & -0.1906+0.315i & 0.0052-0.0422i\\
0.3518+0.0347i & 0.2181-0.0184i & 0.2035+0.2981i & -0.2619-0.1756i & 0.0745+0.2214i & -0.3757-0.2796i & -0.0407+0.1763i & -0.0282+0.5385i\\
0.1854-0.3011i & 0.1739+0.4225i & 0.0297+0.1532i & -0.1691-0.2287i & 0.258+0.129i & 0.0451+0.3754i & -0.073-0.529i & -0.1907-0.1153i\\
-0.1763-0.3065i & -0.2514-0.0616i & -0.2356+0.0748i & -0.6395-0.2149i & -0.4043-0.0311i & 0.1943-0.1172i & 0.2272-0.0731i & 0.1031+0.091i\\
0.2343+0.2647i & -0.5026-0.1922i & 0.4268-0.0851i & -0.1434+0.0711i & -0.0345-0.3266i & 0.0132-0.041i & -0.1625-0.4256i & -0.2028+0.1095i\\
0.2445+0.2554i & 0.1842+0.1208i & 0.2086+0.2099i & -0.1123-0.2286i & -0.1692-0.1362i & 0.1655-0.1419i & -0.2248+0.0226i & 0.5754-0.444i\\
\end{bmatrix}},
\end{equation}

\begin{equation}
V_2 = {\tiny
\begin{bmatrix}
-0.4008+0.299i & -0.2726-0.0633i & -0.5762+0.0519i & -0.3259-0.4804i\\
0.3645+0.3422i & -0.2194+0.2229i & 0.476-0.3105i & -0.0414-0.5723i\\
0.0648-0.4958i & 0.3064+0.0338i & 0.1171+0.564i & -0.0865-0.5618i\\
-0.4988-0.034i & 0.1224+0.8449i & 0.0872-0.0542i & 0.0955+0.039i\\
\end{bmatrix}},
\end{equation}

\begin{equation}
W_2 = {\tiny\begin{bmatrix}
-0.2271+0.271i & -0.1472+0.254i & 0.321+0.1788i & 0.1722-0.0949i & -0.162-0.4895i & -0.1554+0.4099i & 0.1569-0.0484i & -0.3607-0.0085i\\
-0.3475+0.0651i & 0.3698+0.069i & 0.1526-0.0904i & -0.2671+0.0309i & -0.209-0.117i & 0.1731-0.4028i & 0.1465-0.143i & -0.0624-0.5782i\\
-0.1736-0.308i & -0.1436+0.287i & 0.2615-0.184i & -0.2339+0.5177i & -0.0259+0.0096i & 0.1182-0.0133i & -0.3389+0.3749i & -0.2362+0.1446i\\
0.3238+0.1419i & 0.142+0.3402i & -0.2921-0.2162i & -0.0305-0.208i & 0.2057-0.5327i & -0.0957-0.319i & -0.3229+0.1092i & -0.0894+0.0403i\\
0.3455+0.0751i & -0.0961+0.4293i & -0.107+0.2211i & -0.1628+0.5174i & 0.0118+0.015i & -0.0694-0.1127i & 0.3936-0.3791i & 0.0777+0.0661i\\
0.2639-0.2353i & -0.3539+0.0755i & 0.2208-0.5372i & 0.1758-0.1027i & -0.2688+0.035i & -0.38-0.0746i & 0.1909+0.0368i & 0.1329-0.2938i\\
-0.3046-0.1795i & -0.1508+0.2398i & 0.1115+0.2248i & 0.0133-0.3706i & -0.1492+0.1805i & -0.2096-0.5272i & -0.0013-0.159i & -0.0742+0.4322i\\
-0.1061+0.3373i & -0.2949+0.2044i & -0.2965+0.2251i & 0.1974+0.0994i & -0.4443+0.1544i & 0.0274-0.0227i & -0.4356+0.0898i & 0.2576-0.2688i\\
\end{bmatrix}}.
\end{equation}

\begin{figure}
    \includegraphics[width=0.7\textwidth]{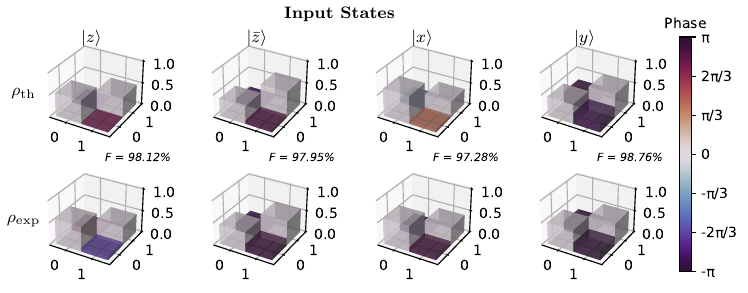}
    \caption{\textbf{Output density matrices of the input state set $\mathcal{B}$ in a general superchannel $\hat{\C S}_g$ scheme.}}
    \label{fig:decom_rho_final}
\end{figure}

\begin{figure}
    \includegraphics[width=0.7\textwidth]{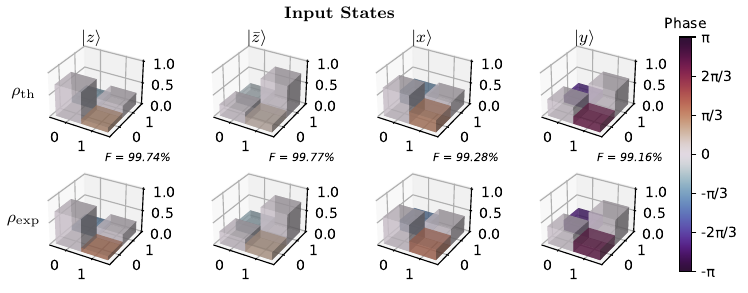}
    \caption{\textbf{Output density matrices of the input state set $\mathcal{B}$ in a decomposed superchannel $\hat{\C S}_g^1$ scheme.}}
    \label{fig:decom_rho_1}
\end{figure}

\begin{figure}
    \includegraphics[width=0.7\textwidth]{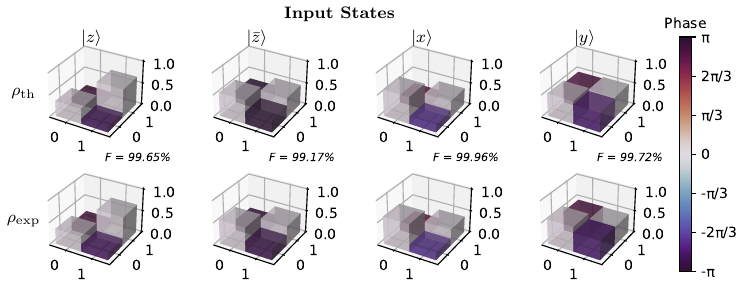}
    \caption{\textbf{Output density matrices of the input state set $\mathcal{B}$ in a decomposed superchannel $\hat{\C S}_g^2$ scheme.}}
    \label{fig:decom_rho_2}
\end{figure}

The output density matrices of the input state bases in the general superchannel $\hat{\C S}_g$ scheme in theory and experiment are presented in Fig.~\ref{fig:decom_rho_final}. The state fidelities between the theoretical and experimental density matrices of $\ket{z}$, $\ket{\bar{z}}$, $\ket{x}$ and $\ket{y}$ under the channel $\C E$ are $98.12\%$, $97.95\%$, $97.28\%$ and $98.76\%$, respectively. While that of its decomposed superchannels $\hat{\C S}_g^1$ and $\hat{\C S}_g^2$ schemes are presented in Fig.~\ref{fig:decom_rho_1} and Fig.~\ref{fig:decom_rho_2}, respectively. The fidelities of that in $\hat{\C S}_g^1$ scheme are $99.74\%$, $99.77\%$, $99.28\%$ and $99.16\%$, and the fidelities of that in $\hat{\C S}_g^2$ scheme are $99.65\%$, $99.17\%$, $99.96\%$ and $99.72\%$.

\section{More examples of quantum superchannel}
\label{app2}


\subsection{Entanglement-assisted quantum communication}

\begin{figure}[b!]
    \centering
    \includegraphics[width=0.6\textwidth]{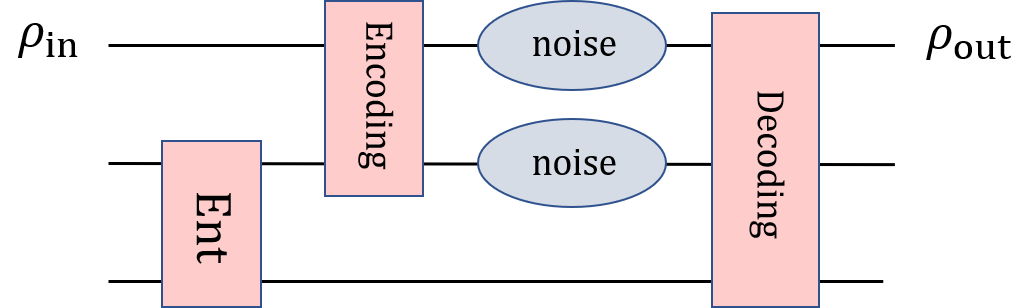}
    \caption{Circuit for entanglement-assisted quantum communication.}
    \label{fig:EA}
\end{figure}

A notable protocol that was developed before the emerge of superchannel theory 
is the entanglement-assisted quantum communication~\cite{entangle-com}. 
A circuit of it is shown in Fig.\ref{fig:EA}, 
where Ent represents the pre-existed entangled states shared by Alice and Bob.
This circuit is of the form of superchannel with noise in the communication process as the input channel. 
The Ent and Encoding operations together form the pre-operation of the superchannel, 
while the decoding operation is the post-operation,
which could consume additional ancillary qubits. 

The entangled state $\ket{\text{Ent}}$ is used as a resource, 
and it is assumed to be free from noise.
For instance, a common resource is the ebit,
which can be generated and distributed via a specific protocol,
then Alice and Bob each needs to store their qubits for the later usages. 
The flying qubits that will subsequently be transmitted
suffer from noise, 
hence requiring quantum error correction. 
As is well known, a large class of quantum codes 
is the entanglement-assisted error correction codes,
which possess some interesting features compared with codes 
without entanglement assistance~\cite{Hsieh,guenda}.

\subsection{Noise-adapted quantum error correction}

Using superchannel theory, here we show a construction of error correction codes
that are noise-adapted. 
A common noisy channel that exists in many experimental systems is the amplitude-damping (AD) channel
defined by two Kraus operators 
\begin{equation}
    K_0=\begin{pmatrix}
       1 & 0  \\
       0 & \sqrt{1-\lambda}
   \end{pmatrix}, \; 
    K_1=\begin{pmatrix}
       0 & \sqrt{\lambda}  \\
       0 & 0
   \end{pmatrix} 
\end{equation}
for $\lambda\in [0,1]$ as the damping parameter that encodes the evolution time. 
Our codes are $\gamma$-dependent and approximate, 
similar with some other codes in literature~\cite{PhysRevA.56.2567,RN426,RN429,4675715,Ng,MN12}.
In particular, we have an optimization algorithm that 
can find a code which can improve the quality of a noisy qubit. 

\begin{figure}
    \centering \includegraphics[width=1\textwidth]{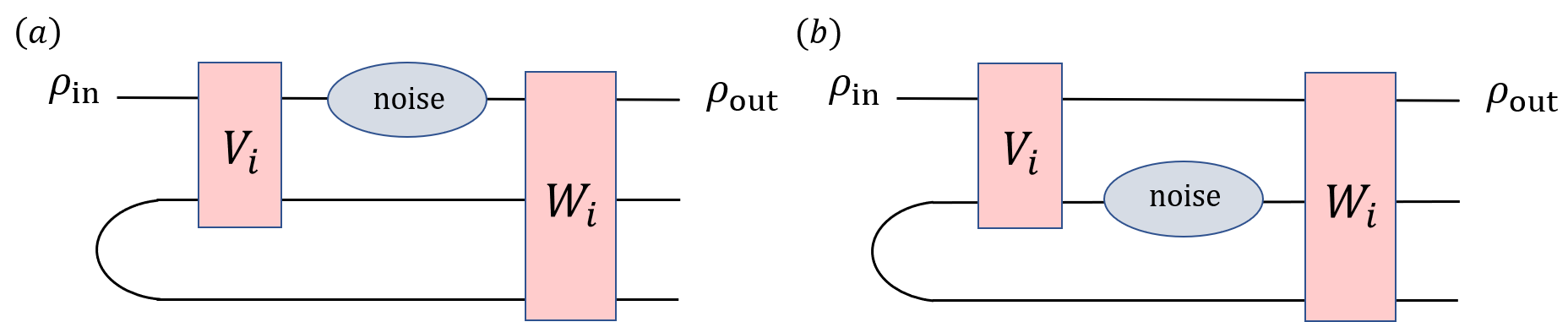}
    \caption{The ebit-assisted communication when AD noise occurs in the 1st qubit (a)~and 2nd qubit (b).}
    \label{adcircuit}
\end{figure}

Adapted to the NMR simulator we have, 
we use one qubit as the ancilla to generate the AD noise, 
and three qubits as the codes. 
We assume only one qubit suffers the AD noise
so that we can treat our codes as distance-3 against the AD noise, 
approximately. 
Under this assumption, 
we theoretically consider three cases as in Fig.~\ref{adcircuit}, 
namely, either the 1st or 2nd qubit is noisy, 
or with equal probability one of them is noisy,
and the 3rd one, as a part of the ebit, is assumed to be noise-free.

\begin{figure}[b!]
    \centering \includegraphics[width=0.5\textwidth]{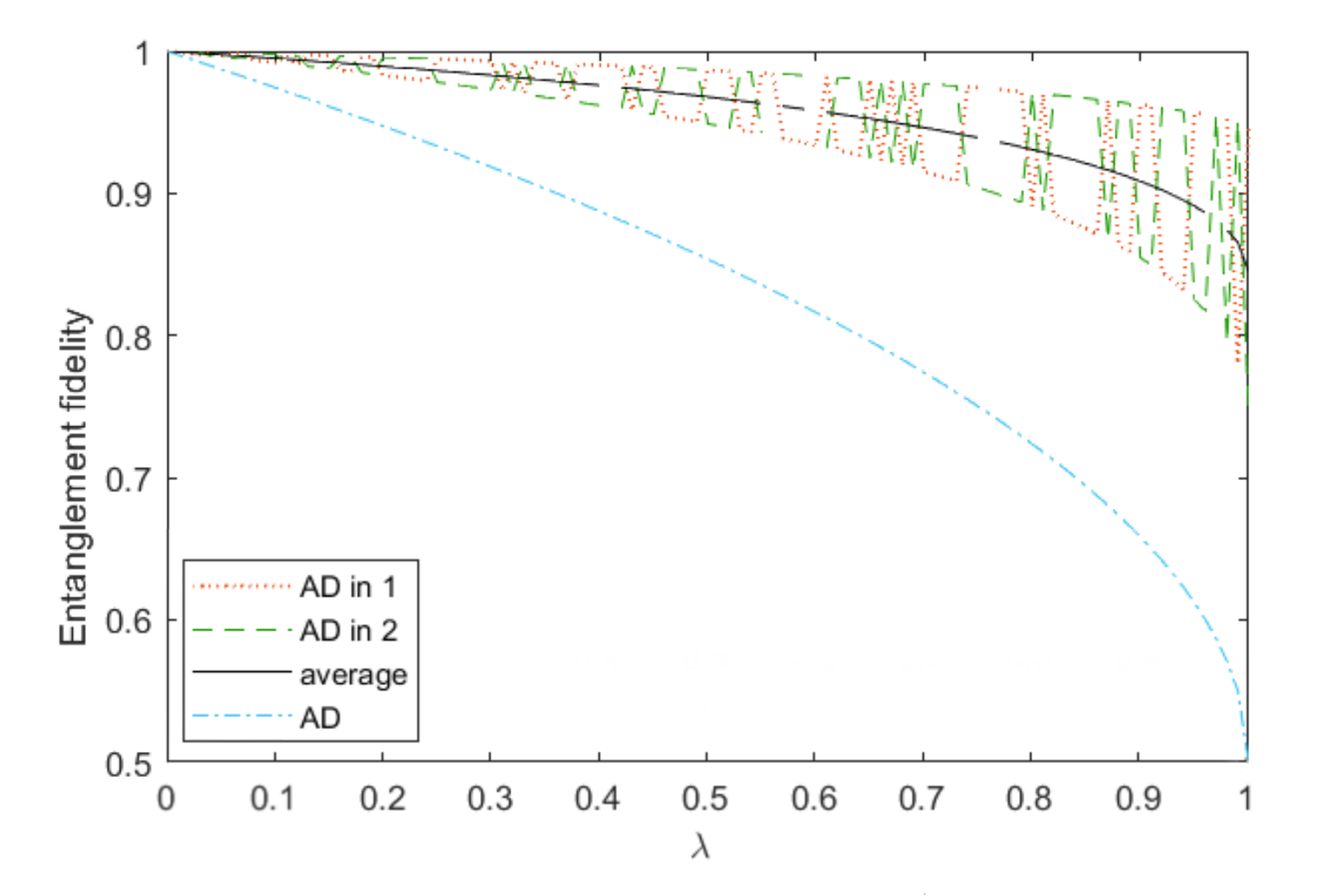}
    \caption{The entanglement fidelity with (black line) and without  (blue dashed line) error correction
     as a function of the damping rate $\lambda$.}
    \label{admatlab}
\end{figure}

In Fig.~\ref{admatlab}, we show the case when either one of the first two qubits is noisy. 
The black dashed line is the optimal result we find numerically,
which is the average of the green and red lines.
The fidelity is the entanglement fidelity between the error-corrected noise 
and a perfect identity channel.
It is obvious to see when the averaged fidelity is optimal, 
the fidelity for merely one noise (green or red) does not need to be so, though.
For each given $\gamma$, a code defined by the pair, encoding $V$ and decoding $W$ operations, 
is found. 
We can see that our codes indeed suppress the AD noise quite well,
and the fidelity is high especially when $\gamma$ is small. 

The primary example above shows that superchannel could be a promising framework
to design error correction codes.
For better and more practical codings to construct real logical qubits, 
we need to consider larger systems,
which is left for future investigation.


\section*{Acknowledgements}

H. L., K. W., and S. W. contributed equally to this work. We acknowledge the National Natural Science Foundation of China under Grants No. 12047503 and 12105343 (K. W., D.-S. W.), 12005015 (S. W.), 11974205 (H. L., S. W., F. Y., X. C., G.-L. L.),  and the National Key Research and Development Program of China (2017YFA0303700), 
The Key Research and Development Program of Guangdong province (2018B030325002), 
Beijing Advanced Innovation Center for Future Chip (ICFC), the Beijing Nova Program (20230484345) (H. L., S. W., F. Y., X. C., G.-L. L.).


\bibliography{ext}{}
\bibliographystyle{unsrt}


\end{document}